\documentclass[fleqn,usenatbib]{mnras}

\usepackage{newtxtext,newtxmath}

\usepackage[T1]{fontenc}

\DeclareRobustCommand{\VAN}[3]{#2}
\let\VANthebibliography\thebibliography
\def\thebibliography{\DeclareRobustCommand{\VAN}[3]{##3}\VANthebibliography}

\usepackage{graphicx}	% Including figure files
\usepackage{amsmath}	% Advanced maths commands
\usepackage{supertabular}

\title{
Pushchino Multibeam Pulsar Search. VIII. Pulsar with a period of 40.9~s in observations of the LPA LPI
}
\author[Tyul'bashev et al.]{S. A. Tyul'bashev, $^{1}$\thanks{E-mail: serg@prao.ru (SAT)}
G. E. Tyul'basheva,$^{2}$
S. A. Andrianov,$^{1}$
M. A. Kitaeva$^{1}$
\\
	$^{1}$ P.N. Lebedev Physical Institute of the Russian Academy of Sciences, Astro Space Center, Pushchino Radio Astronomy Observatory,\\
	Radiotelescopnaya 1a, Moscow reg., Pushchino, 142290, Russia \\
	$^{2}$ Institute of Mathematical Problems of Biology RAS (IMPB RAS), Branch of Keldysh Institute of Applied Mathematics,\\ Russian Academy of Sciences, Pushchino, 142290 Russia
}%

\date{2025}

\pubyear{2025}

\begin{document}
	%\label{firstpage}
	\maketitle
	
\begin{abstract}
A search has been carried out for the pulsar J0311+1402, which has a period of $P = 40.9$ s, in the data archive of the Large Phased Array (LPA) radio telescope. When searching using fast folding algorithm (FFA), periodic pulsar radiation at a frequency of 111 MHz was not detected. In 3321 observation sessions lasting 5 minutes, 35 strong pulses were detected with a signal-to-noise ratio (S/N) greater than 10. Some of the pulses have a complex multi-peak structure consisting of narrow details, while some of the pulses are single-component. The peak flux densities of the details of these strong pulses range from 2 to 11 Jy. The peak value ($S_{\rm p} = 2$\,Jy) and the integral ($S_{\rm i} = 7$\,mJy) flux density in the average profile were obtained from the strong pulses. It is shown that pulsar pulses in the meter-wavelength range arrive sporadically, and the pulsar is similar in its properties to a rotating radio transient (RRAT). The pulsar has the minimal dispersion measure, the minimal distance from the Sun, and the minimal pseudo-luminosity of all known pulsars. Pulsar timing made it possible to improve the previously obtained value of the period ($P$) and to estimate the period derivative ($\dot P$). In the dependency of timing residuals (TRs) from the times of arrival (TOA) of pulses discontinuities are visible, when no pulses were observed. The duration of these breaks can be hundreds of days.

\end{abstract}

\maketitle 

\section {Introduction}

Pulsars were discovered almost 60 years ago (\citeauthor{Hewish1968}, \citeyear{Hewish1968}).  
Nevertheless, a complete understanding of the mechanisms of their radio emission is still lacking.  
The only point beyond doubt is that the radio emission is produced by charged particles in the strong magnetic field of a rapidly rotating neutron star. Despite the absence of a universally accepted emission mechanism, the magneto-dipole model is the one most often used in practice for estimates (see, e.g., the handbook \citeauthor{Lorimer2004} (\citeyear{Lorimer2004})).

While radiating, the neutron star gradually loses energy and spins down; i.e.\ its rotation period, $P$, increases. The spin-down rate is described by the period derivative, $\dot P$.  As rotational period grows, the conditions governing the possibility of radio emission change. Because the initial magnetic fields and initial spin periods of neutron stars may differ by many orders of magnitude, the lifetimes of radio pulsars vary widely.

The continuously updated pulsar catalogue (\citeauthor{Manchester2005}, \citeyear{Manchester2005})\footnote{\url{https://www.atnf.csiro.au/research/pulsar/psrcat/}}, which in May 2025 contains more than 3\,000 radio pulsars, shows that the median period is $P \sim 0.5$ s.The number of pulsars in the catalogue drops sharply with increasing period: fewer than 300 pulsars have $P > 2$ s, fewer than 40 have $P > 5$ s, and only 5 radio pulsars with $P > 10$ s are listed. All the latter discovered since 2018.

Pulsars with $P > 10$ s lie near or beyond the so-called death line on the $P/\dot P$ diagram (see, e.g., Fig.~4 in \citeauthor{Wang2025} (\citeyear{Wang2025})). The magneto-dipole mechanism cannot explain the observed energy of their radiation. In general, the total number of long-period pulsars is expected to be small.

The discovery of pulsars with large periods is often an accident. For example, the pulsar J0250$+$5854 with $P = 23.5$ s was found in a survey designed to detect pulsars with periods up to 16 s (\citeauthor{Tan2018}, \citeyear{Tan2018}). For the search, Fast Fourier-transform (FFT) folding algorithms were used. At low frequencies (long periods), however, low-frequency (red) noise is present in the power spectra, limiting the detection of long-period pulsars. The pulsar J0901$-$4046 with $P = 75.8$ s was accidentally detected through its single-pulse emission during observations of the object Vela~X--1 (\citeauthor{Caleb2022}, \citeyear{Caleb2022}). The pulsar J0311$+$1402 with $P = 40.9$ s was discovered via its single-pulse emission in the decimeter wavelength range during equipment tests for interplanetary scintillation observations (\citeauthor{Wang2025}, \citeyear{Wang2025}).

The total number of long-period pulsars appears to be small. For instance, on the Large Phased Array radio telescope (LPA) of the Lebedev Physical Institute (LPI) at 111 MHz, a dedicated survey covering an area of 6\,300 square degrees ($+21^{\circ} < \delta < +42^{\circ}$) searched for pulsars with periods $2 < P < 90$ s using a five-year data set. With a sensitivity of 1--3~mJy outside the Galactic plane, not a single pulsar with $P > 10$ s was detected (\citeauthor{Tyulbashev2024}, \citeyear{Tyulbashev2024}). Fast-Folding-algorithm (FFA) summed periodograms were employed. The maximum period found in this survey, $P = 7.334$~s, belongs to the discovered pulsar J1951$+$28.

In the present work we report on searches for both periodic and single-pulse emission from the pulsar J0311+1402 in the meter-wavelength range. In the following sections we discuss the search, present refined rotational parameters derived from timing a ten-year data set, and discuss the properties of the pulsar.

\section{Search for the pulsar}

To search for the pulsar J0311+1402 we used data obtained with the LPA radio telescope. The  telescope has two independent beam patterns (LPA1 and LPA3). The steerable pattern (LPA1) is intended for pulsar studies. The second pattern (LPA3) is stationary: the beam directions are fixed in elevation and lie in the plane of celestial meridian. LPA3 consists of 128 beams that were connected to recorders at different times. Recording is carried out simultaneously in two frequency-time modes: sampling rates of 80~Hz and 10~Hz (12.5~ms and 100~ms per sample) with channel widths of 78~kHz and 415~kHz, which have remained unchanged since the 2013 antenna upgrade. The total bandwidth, 2.5~MHz, split into 32 and 6 frequency channels respectively (\citeauthor{Shishov2016}, \citeyear{Shishov2016}, \citeauthor{Tyulbashev2016}, \citeyear{Tyulbashev2016}). The beam size is approximately $0.5^\circ \times 1^\circ$. The high instantaneous sensitivity of LPA3 is due to its large effective area ($A_{\rm eff}\approx 45\,000$~m$^2$). A survey for pulsars and transients covering almost half of the celestial sphere has been carried out continuously, starting in August 2014. The accumulated data volume is about 450~TB.

The pulsar J0311$+$1402 lies in the declination range that is monitored daily within the "Pushchino multibeam pulsar search" (PUMPS) survey (\citeauthor{Tyulbashev2022}, \citeyear{Tyulbashev2022}). Processing for the pulsar search with periods 2~s $< P <$ 90~s using FFA spectra was performed for declinations $-9^{\circ} < \delta < +42^{\circ}$. A ten-year observation interval (paper in preparation) was used. From the processed data we extracted the strip containing the pulsar coordinates to search for periodic emission.

Indirect indications suggested that periodic emission should be detected. Indeed, the expected sensitivity achieved after summing FFA spectra is no worse than 1--3~mJy outside the Galactic plane at zenith. The expected peak flux density ($S_{\rm p}$) in the pulsar's average profile, scaled from 816~MHz to 111~MHz assuming a spectral index $\alpha = 2.3$ ($S\propto\nu^{-\alpha}$) (\citeauthor{Wang2025}, \citeyear{Wang2025}), should be 3~Jy. The expected integral flux density ($S_{\rm i}$) at 111~MHz, assuming a pulse width of 450~ms and a triangular shape, can be estimated as 16~mJy. Thus, if the pulsar were a canonical slow pulsar emitting a pulse on every rotation, it would certainly have been detected in the processed data. However, the pulsar was not found in the FFA spectra.

In addition to the search for periodic emission, we carried out a search for single pulses from the pulsar. The pulses are broad: according to \citeauthor{Wang2025} (\citeyear{Wang2025}) their width is about 450~ms. Consequently, the slope of the dispersion-delay line across the LPA bandwidth may be weak and hard to see by eye. In this case the only reliable criterion in the pulse search is the known pulsar period. The LPA data archive contains more than ten years of observations, each sky position in the declination range $-9^{\circ} < \delta < +42^{\circ}$ has been observed for about 200~hours. If the raw-data plots are retained, a visual pulse search can be performed. After detecting at least two ''suspicious'' pulses in a plot, the arrival times can be estimated and the difference between them checked against the known pulsar period. The pulse width is large enough that the search can be carried out in low time-resolution data. In these data the sampling interval is 0.1~s, which automatically averages the pulse and increases.

We took the 2.5~MHz bandwidth data recorded between 01~January~2015 and 28~April~2025 (more than 3700 sessions) and extracted 4-minute segments. Plots were produced for visual pulse searching. Over 240~s, depending on the pulse phase, five or six pulses may appear. Most of the visible signals are ordinary pulse interferences.

\begin{figure}
	\includegraphics[width=0.9\columnwidth]{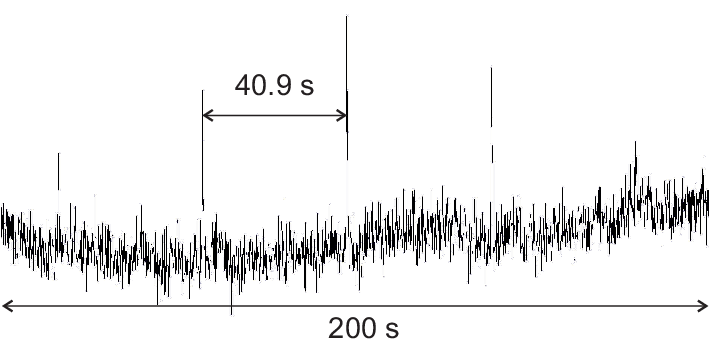}
	\caption{The figure shows 200~s of low time-frequency-resolution data recordings for 06 February 2025. The data have no additional processing. There is no DM compensation, time averaging and noise interference filtering.}
	\label{Fig1}
\end{figure}

In manual mode we found 145 sessions (days) in which 2 to 5 pulses were detected with separations that are multiples of $P = 40.9$~s. Fig.~\ref{Fig1} shows raw low-resolution data for one of the best days. Five pulses are visible in the plot. In total, 6 sessions were found in which 5 pulses were observed.

\section{Dispersion measure of pulsar}

\begin{figure}
\includegraphics[width=0.9\columnwidth]{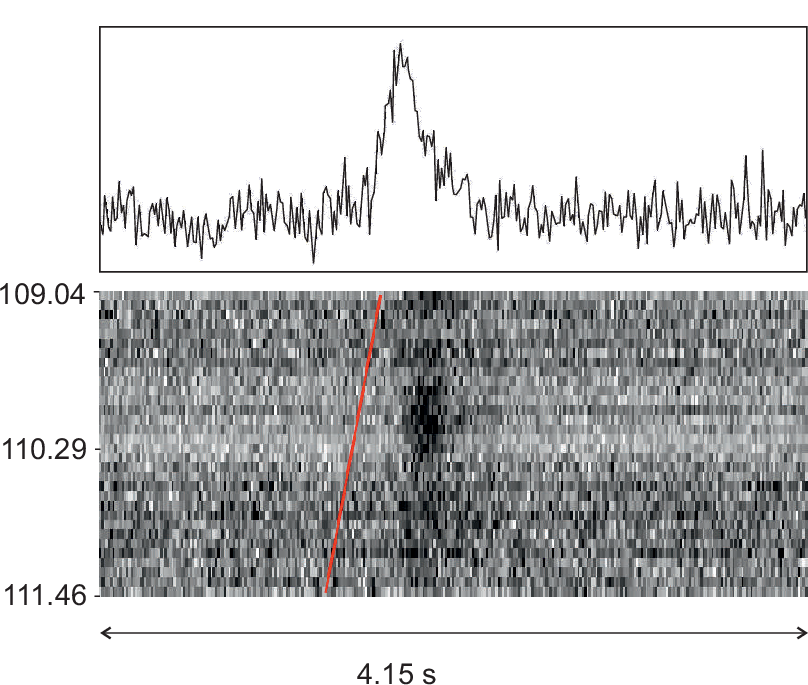}
\caption{The upper panel shows the pulse profile, that is averaged with DM$=19.9$~pc~cm$^{-3}$. The bottom panel shows the dynamic spectrum of this pulse. The red line represents expected slope with DM$=20$~pc~cm$^{-3}$. The vertical axes represent relative intensity and observational frequency respectively. The horizontal axes represent time.}
\label{Fig2}
\end{figure}

According to \citeauthor{Wang2025} (\citeyear{Wang2025}), the pulsar J0311+1402 has a small dispersion measure, DM = 19.9~pc~cm$^{-3}$. However, inspection of the pulses in high time-resolution data revealed that on the dynamic spectra the visible slope of the pulse does not match the expected slope (see the red line in Fig.~\ref{Fig2}). That is, from the appearance of the pulse's dynamic spectrum it is clear that the DM is smaller than 19.9~pc~cm$^{-3}$.

\begin{figure}
	\includegraphics[width=0.9\columnwidth]{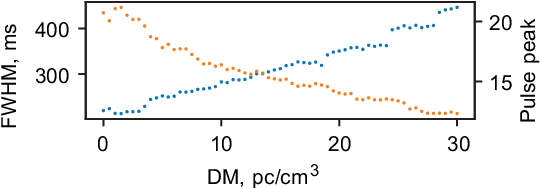}
	\caption{The estimation of DM. The blue dots (left vertical axis) shows estimations of averaged profiles half-widths in ms. The orange dots (right vertical axis) shows amplitude of the profiles in conventional units. The horizontal axis represents DM's grid.}
	\label{Fig3}
\end{figure}

To estimate DM we used two methods. In the first method, using the timing solution (see the next section) we produced an average profile by summing the 35 strongest pulses. The strongest pulses are those whose S/N exceeds 10 after averaging the signal over 8 sample points. The average profile was obtained by stepping DM from 0 to 30~pc~cm$^{-3}$ in steps of 0.5~pc~cm$^{-3}$. Thus, we generated 60 average profiles corresponding to different DMs. For each profile we measured its amplitude and its full width at half-maximum (FWHM). Fig.~\ref{Fig3} plots the amplitude and FWHM values of the average profiles. It is evident that the amplitude of the average profile peaks at DM = 1--1.5~pc~cm$^{-3}$, while the FWHM reaches a minimum in the same DM interval.

\begin{figure}
	\includegraphics[width=0.7\columnwidth]{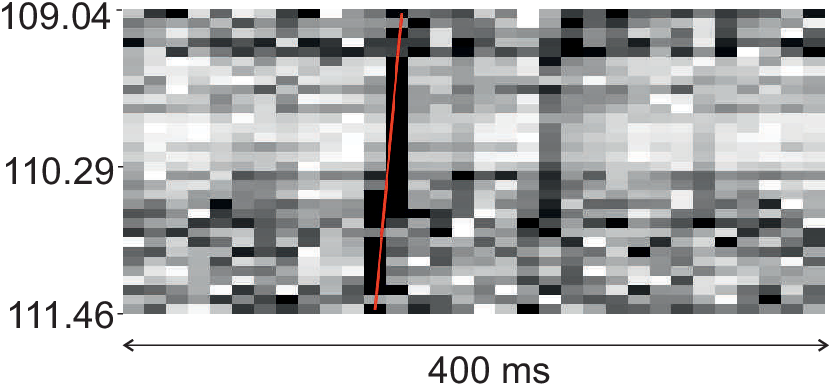}
	\caption{Dynamic spectrum of the brightest of the detected pulses, recorded on 06 April 2022. The observation frequencies are placed along the vertical axis. The horizontal axis represents time. The red line, aligned with the strongest narrow detail in the pulse, shows the slope expected with DM$=1$~pc~cm$^{-3}$.}
	\label{Fig4}
\end{figure}

In the second method we estimated DM from the narrow features of strong pulses. When examining strong pulses, some were found to contain narrow structures (multiple peaks) lasting only one sample (see the dynamic spectrum in Fig.~\ref{Fig4}). The visible drift across the frequency band also corresponds to one sample, which is 12.5~ms. The total bandwidth is 2.5~MHz. It is easy to estimate that a one-sample drift in the dynamic spectrum corresponds to DM $\simeq$ 1~pc~cm$^{-3}$.

Our final estimate of the pulsar's dispersion measure is DM = $1\pm0.5$~pc~cm$^{-3}$. This value is many times smaller than that given in \citeauthor{Wang2025} (\citeyear{Wang2025}). The final timing (see the next section) was performed assuming DM = 1~pc~cm$^{-3}$.

\section{Timing of J0311+1402}

Timing of the pulsar J0311+1402 was carried out in several iterations. First we attempted to locate visually the pulses found in low-resolution data within the high-resolution data. Out of about 300 pulses visible in the low-resolution data, only 62 remained that could be identified by eye in the high-resolution data. Thus, we selected the strongest pulses, and the initial timing was based on them.

The LPA3 recorders receive the data-recording clock signal from quartz oscillators. The accuracy of determining pulse times of arrival  (TOAs) is low, so pulsar timing from such TOAs is non-trivial. The procedure for timing data with poor time tagging is described in detail in \citeauthor{Andrianov2025} (\citeyear{Andrianov2025}). Its essence is that during a single observation session an accuracy of 10$^{-5}$~s in the time tags is sufficient to build the average profile, and this accuracy is provided by the quartz oscillator. To maintain correct pulse phase across sessions (different observing days) a pulsar timescale was created, using bright pulsars with well-known parameters. As a result, time corrections for the quartz oscillator were obtained for any instant. This two-step procedure allows the pulsar period to be determined with an accuracy of $\sim$10$^{-9}$~s over a ten-year data span.

According to \citeauthor{Wang2025} (\citeyear{Wang2025}), a six-month data set for J0311+1402 containing 20 TOAs yielded to a period accuracy of 2$\cdot$10$^{-7}$~s. This accuracy corresponds to pulse-phase loss on time scales of tens of years. For the LPA3 timing we adopted the rotational parameters and coordinates from \citeauthor{Wang2025} (\citeyear{Wang2025}) as starting values and used TOAs of the 62 pulses spread over 10~years. The timing solution was obtained with the program TEMPO2 (\citeauthor{Hobbs2006}, \citeyear{Hobbs2006}) and with special software that enables TOAs determined from quartz clocks to be used (\citeauthor{Andrianov2025}, \citeyear{Andrianov2025}).

After obtaining timing residuals (TRs) and extracting the rotational parameters via the strong pulses, we searched for pulses in all observing sessions. To find weak pulses invisible in the raw data, a running average with an 8-sample window was applied in the region where a pulse was expected. The averaging step (100~ms) was chosen according to the expected pulse width of 200--300~ms. The averaging increases S/N of the pulses and enables weak pulses to be revealed. In ordinary 32-channel observations without averaging, a pulse with S/N = 7 (a reliably detected pulse) corresponds to a peak flux density $S_{\rm p} = 2.1$~Jy (\citeauthor{Tyulbashev2018}, \citeyear{Tyulbashev2018}). After 8-sample averaging we can detect pulses about three times weaker, down to $S_{\rm p} = 0.7$~Jy. 

Taking into account that the search is conducted at pre-computed pulse phases, the S/N threshold was lowered from 7 to 4, corresponding to $S_{\rm p} = 0.4$~Jy. When searching for weak pulses we also required that the deviation from the expected (pre-computed) pulse location be no more than 0.5~s. 

As the dispersion measure is very small, we cannot distinguish a pulse from impulsive interference by eye. A dedicated study showed that the total number of pulse bursts in the LPA3 data is small: on average, 10 bursts occur per hour in an LPA3 beam (\citeauthor{Samodurov2023}, \citeyear{Samodurov2023}). The probability of a burst falling within the pre-computed pulse window is $10/3600 = 0.0028$. Nevertheless, accidental coincidence of a burst with the window cannot be completely excluded. Therefore, we carried out an additional check, verifying whether burst-like signals coincident in time with the found pulses are present in beams adjacent (above and below) to the central beam, containing the pulsar.

As a result of the pulse search with the pre-computed phase and with rejection of interference profiles, in 3486 sessions 35 pulses, visible in the raw data without averaging and 180 pulses visible in the averaged data were found.
The final sample comprised:
\begin{itemize}
\item 62 pulses found by eye in the high time-resolution data;

\item 35 pulses found by the timing criterion in the high time-resolution data;

\item 180 pulses found by the timing criterion in the 8-sample averaged data.

\end{itemize}

For the total 97 pulses, found in the high-resolution data,
TRs were re-defined and pulsar rotational parameters were refined. For the detected pulses we estimated the peak flux densities of the narrow features in the strongest and weakest pulses: $S_{\rm p}=11$~Jy and 2~Jy, respectively.

Because of the possible coincidence of interference with the pre-computed pulse window, the final pulse selection is not obvious. The wider the pre-computed window, the more pulses are registered, and the more of them may be interference. If the search criteria are relaxed, the number of pulses increases to about 1\,000.

\begin{figure}
	\includegraphics[width=0.9\columnwidth]{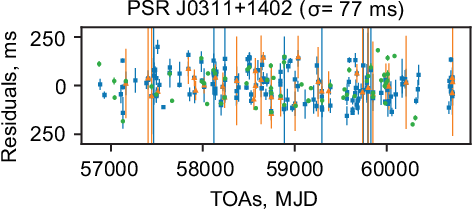}
	\caption{The vertical axis represents timing residuals of the pulsar in ms. The horizontal line represent pulses' times of arrival in Modified Julian Date (MJD). The green circles show TRs of the pulses found in manual search (62 pulses), the orange triangles show TRs of pulses found in the automatic search in high time-resolution data (35 pulses), the blue squares show TRs of pulses found in the automatic search in 8-sample averaged data (180 pulses). The vertical lines show errors of TOAs determining.}
	\label{Fig5}
\end{figure}

After obtaining $P$ and $\dot P$, we built the average profile from the 35 strongest pulses, measured its height (S/N = 85) and its width at half-maximum ($W_{0.5}=212$~ms) (Fig.~\ref{Fig5}). These values were used to estimate the peak and integral flux densities in the average profile: $S_{\rm p}=2$~Jy and $S_{\rm i}=7$~mJy. These flux estimates are not canonical. For a canonical estimate of the integral flux density it is necessary to sum all pulses regardless of whether they are detected.

When 3201 sessions are summed with timing solution (assuming DM = 0~pc~cm$^{-3}$) the resulting average profile has S/N = 20, substantially lower than the S/N = 85 obtained from the 35 pulses. Moreover, the average profile from all sessions may contain spurious pulses that increase the apparent flux density. Therefore, the average profile from all sessions was used only to obtain upper limits for the peak ($S_{\rm p1}\le 140$~mJy) and integral ($S_{\rm i1}\le 0.7$~mJy) flux densities. The width at half-maximum of the average profile from all sessions is $\sim400$~ms (Fig.~\ref{Fig6}).

According to \citeauthor{Wang2025} (\citeyear{Wang2025}), the peak flux density in the average profile is $S_{\rm p}=30$~mJy. Using our estimate $S_{\rm p}=2$~Jy from the strong pulses, we obtain an upper limit for the spectral index $\alpha$ between 111 and 816~MHz: $\alpha\le 2.1$. This spectral index is practically identical to the earlier estimate $\alpha=2.3\pm0.2$ obtained in \citeauthor{Wang2025} (\citeyear{Wang2025}) between 816 and 1284~MHz.

Fig.~\ref{Fig5} presents the timing residuals of J0311$+$1402 over a 10-year span. The strong pulses found in the raw data without additional averaging (orange triangles and green circles) were used for timing. As mentioned above, an additional 180 pulses were found after averaging. Their residuals (blue squares) are also plotted in Fig.~\ref{Fig5} but were not used in the timing. The good overlap between the orange, green and blue regions confirms that the weak-pulse selection is correct. Gaps, lasting from a hundred to several hundreds days, are visible in the residuals, when no pulses were registered.

Below we list the rotational parameter and coordinate estimates for J0311$+$1402 referred to epoch MJD~51544:

%\begin{array}{rcl}
$P_0  = 40.910690842 \pm 6.0\times10^{-8}$~s,\\[2pt]
$ \dot P  =  7.871\times10^{-15} \pm 9.5\times10^{-17}$~s/s,\\[2pt]
$F_0  =  0.024443488472 \pm 3.6\times10^{-11}$~1/s,\\[2pt]
$F_1  =  -4.703\times10^{-18} \pm 5.7\times10^{-20}~1/s^{2}$,\\[2pt]
$DM  =  1 \pm 0.5~$~pc cm$^{-3}$
%\end{array}\\
The timing span is 10.5~years.  \\
The number of pulses used in the timing is 97 (62 + 35).

Compared with the earlier timing \citeauthor{Wang2025} (\citeyear{Wang2025}), we improved the period by several times and obtained a value of $\dot P$ instead of the previous upper limit. Having estimates of $P$ and $\dot P$, the physical parameters of the pulsar can be estimated under the assumption of magneto-dipole emission (\citeauthor{Lorimer2004} (\citeyear{Lorimer2004}), handbook): characteristic age $\tau_{\rm c}=85$~Myr, surface magnetic field $B_{\rm s}=1.8\times10^{13}$~G, light-cylinder magnetic field $B_{\rm lc}=2.3\times10^{-3}$~G.

We also estimated the distance to the pulsar using the YMW16 model (\citeauthor{Yao2017}, \citeyear{Yao2017}): $d_1=91.4$~pc, and the NE2001 model (\citeauthor{Cordes2002}, \citeyear{Cordes2002}): $d_1=96.9$~pc, via the on-line calculator (\citeauthor{Crawford2022}, \citeyear{Crawford2022})\footnote{\url{https://pulsar.cgca-hub.org/compute}}. The pulsar J0311$+$1402 turns out to be one of the closest known pulsars. Its pseudo-luminosity, using the mean distance from the two models ($d=94$~pc), is $L_{111}=0.06$~mJy~kpc$^{2}$. If the flux density is scaled to 400~MHz with spectral index $\alpha=2.3$, we obtain $L_{400}=0.003$~mJy~kpc$^{2}$, an order of magnitude lower than the smallest pseudo-luminosity value listed in the ATNF catalogue.

\section{Properties of J0311+1402}

\begin{figure}
\includegraphics[width=0.9\columnwidth]{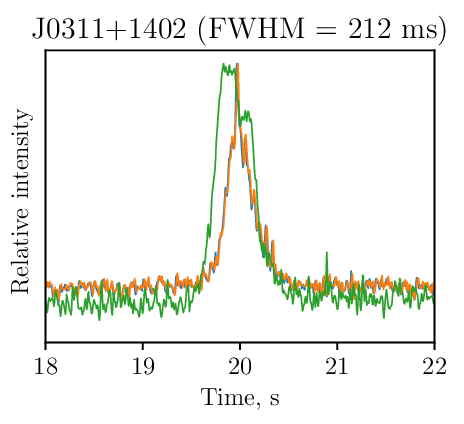}
\caption{Average profile of the pulsar J0311+1402. The blue curve shows the profile obtained from the strongest pulses at DM = 0~pc~cm$^{-3}$, the orange curve shows the profile from the strongest pulses at DM = 1~pc~cm$^{-3}$, and the green curve shows the profile obtained by summing all ten-year observations. The blue curve is almost indistinguishable under the orange line, once again indicating the very small DM of the pulsar. The horizontal axis represent time in seconds from the start of the profile. The profile peak is centered at 20~s. The vertical axis is intensity in arbitrary units.}
\label{Fig6}
\end{figure}

The individual pulses observed from the pulsar exhibit a simple single-component shape, however some pulses consist of many narrow peaks. A number of pulses show asymmetry between their leading and trailing edges. Using the derived values of $P$ and $\dot P$, we summed the 35 strongest detected pulses and obtained the average pulse profile. Fig.~\ref{Fig6} shows a 4-s segment of this average profile. The profile has a simple single-component structure.

The obtained profile is much narrower than that determined in \citeauthor{Wang2025} (\citeyear{Wang2025}), clearly because of the incorrect DM used in that work. The width at half-maximum of the profile (FWHM) is $W_{0.5}=212$~ms when derived from the 35 strongest pulses, and $W_{0.5}=415$~ms when the signal is summed over all sessions after interference rejection. In \citeauthor{Wang2025} (\citeyear{Wang2025}) the individual pulses were reported to be two-component, possibly three-component, whereas our average profile, built from the 35 strongest pulses, appears single-component.

Over the 10-year data span we found fewer than one percent of all pulses that fall within the 5-minute interval when the pulsar crosses the meridian. However, since the sensitivity drops by a factor of 2.5 at the edges of the half-power beam, some weak pulses may be missed. Therefore, for nulling estimates we used the central 2~minutes of meridian transit, where the flux-density correction is less than 15\% of the measured peak. Using only the central part of the beam and pulses detected after 8-sample averaging, the nulling fraction is 0.995. If most of the $\sim$1000 pulses obtained with relaxed search criteria are real, the nulling fraction decreases to 0.98. In both cases the vast majority of pulses are absent, so nulling is the obvious reason why the pulsar was not detected in the FFA search.

FWHMs of the 180 pulses found with 8-sample averaging vary widely: the minimum is 120~ms, the maximum 600~ms, the median 343~ms, and the mean $358\pm105$~ms. This mean FWHM from weak pulses does not match the FWHM of the average profile derived from the 35 strongest pulses.

\section{Discussion and conclusion}

In our observations there are sessions when every pulse of the pulsar that enters the LPA3 beam is visible; however, as noted above, in the overwhelming majority of sessions not a single pulse is seen visually. In Fig.~2 of \citeauthor{Wang2025} (\citeyear{Wang2025}) raw data from two MeerKAT sessions, each 39~min long, are shown. At 816~MHz almost every pulse of the pulsar is visible in both sessions. The same paper reports the detection of the pulsar's pulses with ASKAP and GBT, noting that pulses are detected not in every session. The authors suggest that their absence may be due to interstellar scintillation, radio-frequency interference, or other causes. Nulling as a reason for the absence of emission is not mentioned.

Observations of regular emission (see \citeauthor{Wang2025} (\citeyear{Wang2025}) and the present work) give a contradictory picture. In the LPA3 pulsar search for long periods ($P>2$~s) at 111~MHz, J0311+1402 was not detected. With MeerKAT at 816~MHz the periodic emission was found. In some ASKAP observations at 887.5~MHz individual pulses were not detected, but a weak ($\text{SNR}\approx5$) average profile was obtained in one hour exposure. At Parkes, 2368~MHz, weak periodic emission was registered in one out of five sessions, each about 1.5~h long. No regular or single-pulse emission from the pulsar was found in archival ASKAP, MWA, or VLSS data.

Summarizing the observations from \citeauthor{Wang2025} (\citeyear{Wang2025}) and the present work, we conclude that at high frequencies (decimeter wavelengths) the emission of J0311+1402 often disappears, but when it is registered, the pulsar behaves like an ordinary slow pulsar. The LPA sessions are short, and in the vast majority of them not a single pulse is observed. All pulses of the pulsar are visible in only 6 out of more than 3000 sessions. Thus, the pulsar behaves differently at high and low frequencies: at high frequencies it resembles an intermittent pulsar, whereas at low frequencies it looks like an RRAT or a pulsar with a very large nulling fraction. The average profile, obtained by summing all 3201 sessions by the timing solution, indicates not regular emission, but merely the presence of numerous relatively strong pulses in our archival data.

In conclusion, we present the main results obtained from the 111~MHz observations:
\begin{itemize}
\item The pulsar J0311+1402 emit sporadic strong pulses and no visible underlying weak-pulse emission. It resembles an RRAT;

\item The timing allowed us to refine the pulsar period and to obtain an estimate of the period derivative. The derived $\dot P$ is typical for ordinary slow pulsars;

\item The pulsar's magnetic field $B_{\text{s}}=1.8\times10^{13}$~G and characteristic age $\tau_{\text{c}}=85$~Myr, derived assuming magneto-dipole braking, are characteristic for slow pulsars;

\item The spectral index $\alpha=2.1$ estimated between 111 and 816~MHz is practically identical to the earlier value $\alpha=2.3$ obtained between 816 and 1284~MHz \citeauthor{Wang2025} (\citeyear{Wang2025}), and is typical for pulsars;

\item The pseudo-luminosity estimated from the detected pulses is extremely low;

\item A new dispersion measure DM = 1~$\pm$~0.5~pc~cm$^{-3}$ has been obtained. It is significantly smaller than the previous estimate DM = 19.9~pc~cm$^{-3}$ (\citeauthor{Wang2025}, \citeyear{Wang2025}). The pulsar turned out to be one of the closest known to date.

\item The pulse is narrow relative to the pulsar period. The fractional width from our observations is $W_{0.5}/P=0.006$. It corresponds to an emission cone width of 2.2$^{\circ}$.

\end{itemize}

\section*{Acknowledgements}

The authors are grateful to engineers and technicians, maintaining the LPA work and performing its repair. We thank S.~B.~Popov for the preliminary reading of this manuscript and a series of useful suggestions. SAT and MAK appreciate Russian Science Foundation (grant 22-12-00236-$\Pi$; {https://rscf.ru/project/22-12-00236-$\Pi$/}) for financial support of this work.

\end{document}